\documentstyle[pra,aps]{revtex}
\begin{document}
\draft
 
\title{Generalized Heat Kernel Coefficients for a New Asymptotic 
       Expansion}
\author{Alexander A. Osipov\thanks{On leave from the Laboratory of 
        Nuclear Problems, JINR, 141980 Dubna, Russia}, and Brigitte Hiller}
\address{Centro de F\'{\i}sica Te\'{o}rica, Departamento de F\'{\i}sica
         da Universidade de Coimbra, 3004-516 Coimbra, Portugal}
\date{\today}
\maketitle
%%%%%%%%%%%%%%%%%%%%%%%%%%%%%%%%%%%%%%%%%%%%%%%%%%%%%%%%%%%%%%%%%%%%%%%%%

\begin{abstract}
%%%%%%%%%%%%%%%%%%%%%%%
  The method which allows for asymptotic expansion of the one-loop
  effective action $W=\ln\det A$ is formulated. The positively
  defined elliptic operator $A=U+{\cal M}^2$ depends on the external
  classical fields taking values in the Lie algebra of the internal
  symmetry group $G$. Unlike the standard method of Schwinger --
  DeWitt, the more general case with the nongenerate mass matrix
  ${\cal M}=\mbox{diag}(m_1,m_2,\ldots )$ is considered. The first
  coefficients of the new asymptotic series are calculated and their
  relationship with the Seeley -- DeWitt coefficients is clarified. 
%%%%%%%%%%%%%%%%%%%%%%%
\end{abstract}

%%%%%%%%%%%%%%%%%%%%%%%%%%%%%%%%%%%%%%%%%%%%%%%%%%%%%%%%%%%%%%%%%%%%%

\section{Introduction}

%%%%%%%%%%%%%%%%%%%%%%%%%%%%%%%%%%%%%%%%%%%%%%%%%%%%%%%%%%%%%%%%%%%%%
The method of Schwinger -- DeWitt \cite{Schwinger:1951,DeWitt:1965}
allows for calculations of radiative corrections in the coordinate 
space. It can be used in any field theory but reveals to be an 
extremely powerful tool when explicit covariance of calculations 
is needed at all intermediate steps. Gauge theory \cite{DeWitt:1965}, 
quantum gravity \cite{DeWitt:1975}, chiral field theory 
\cite{Ball:1989} belong to those cases. In these and in many other 
cases it is necessary to evaluate the determinant of the positively 
defined elliptic operator which describes quadratic fluctuations of 
quantum fields in the presence of some background field. It contains
in compact form the whole information about the one-loop contribution 
of quantum fields. The proper time formalism of Schwinger solves this 
problem. The result is an asymptotic expansion of the effective action 
in powers of proper time with Seeley -- DeWitt coefficients $a_n$ 
\cite{DeWitt:1965,Seeley:1967}, which accumulate the whole dependence
on background fields. It is remarkable that every term of the
expansion is invariant with respect to transformations of the internal 
symmetry group. This is a consequence of the general covariance 
inherent to the formalism. At the present time the asymptotic
coefficients $a_n$ are well known up to and including $n=5$ for a
general operator of Laplace type. The details can be found, for
instance, in \cite{Ball:1989,Gilkey:1994}. 

In the case of massive quantum fields with a degenerate mass
matrix ${\cal M}=\mbox{diag}(m,m,\ldots )$, it is not difficult to
derive from the proper time expansion an expansion in inverse
powers of $m^2$, since the mass dependence is easily factorized and 
a subsequent integration over proper time leads to the desired result.
The resulting asymptotic coefficients remain the same. If the mass
matrix is not degenerate, i.e. ${\cal M}=\mbox{diag}(m_1,m_2,\ldots )$, 
than its total factorization is impossible because of the 
noncommutativity of ${\cal M}$ with the rest of the elliptic
operator. At the same time a naive factorization by parts breaks the 
covariant character of the asymptotic series. The natural question 
arises: is there any simple way to follow which leads to
factorization, conserving at the same time the explicit covariance 
of the expansion? Fortunately, there is such a method 
\cite{Osipov:2001}. It leads to the generalization of the 
Seeley -- DeWitt coefficients.

%%%%%%%%%%%%%%%%%%%%%%%%%%%%%%%%%%%%%%%%%%%%%%%%%%%%%%%%%%%%%%%%%

\section{Integral representation for the one-loop contribution}

%%%%%%%%%%%%%%%%%%%%%%%%%%%%%%%%%%%%%%%%%%%%%%%%%%%%%%%%%%%%%%%%%
The logarithm of a formal determinant describes the low order
radiative corrections to the classical theory. Let fermion fields
play the role of virtual quanta producing those corrections and
the scalar and pseudoscalar mesons be external background fields.
In this case the real part of the corresponding effective action
can be represented as a proper time integral
\begin{equation}
\label{logdet}  
  W[Y]=-\ln |\det D|=\frac{1}{2}\int^\infty_0\frac{dt}{t}\rho 
       (t,\Lambda^2 )\mbox{Tr}\left(e^{-tD^\dagger D}\right).
\end{equation}
Here the Dirac operator $D$ depends on the background fields, which
are collected in the hermitian second order differential elliptic 
operator $D^\dagger D={\cal M}^2+B=-\partial^2+Y+{\cal M}^2$
in the term $Y$. We use the euclidian metric to define the
effective action $W[Y]$. To make the integral over $t$ 
convergent at the lower limit, the regulator $\rho (t,\Lambda^2)$ 
has been added in (\ref{logdet}). All our conclusions are independent 
of the form of this function which includes the ultraviolet cut off
$\Lambda$.  

Let us assume that we are dealing with chiral field theory, which 
possesses the global  $U(N_f)_L\times U(N_f)_R$ symmetry if the 
fermion fields are massless. The typical example is quantum 
chromodynamics. It is known that the vacuum state of low energy QCD 
is noninvariant with respect to the action of the chiral group and 
the whole system makes a phase transition to the state with massive 
quarks. If one takes into account that the explicit chiral symmetry 
breaking takes place in QCD through the mass terms of current quarks 
one can conclude that not equal current quark masses will lead also 
to not equal constituent quark masses. In order to study this
system at large distances we will need the expansion of the 
effective action in inverse powers of the non degenerate mass 
matrix
\begin{equation}
\label{mass}
  {\cal M}=\sum_{i=1}^{N_f} {\cal M}_iE_i,\quad 
  (E_i)_{jk}=\delta_{ij}\delta_{ik},
  \quad E_iE_j=\delta_{ij}E_j.
\end{equation}
The orthonormal basis $E_i$ belongs to the flavor space where the
chiral group acts on quarks and background fields in accordance with
their transformation properties.

The heat kernel $\mbox{Tr}[\exp (-tD^\dagger D)]$ can be represented
\cite{Schwinger:1951} as a matrix element of an operator acting in the
abstract (unphysical) Hilbert space:
\begin{equation}   
   \mbox{Tr}\left(e^{-tD^\dagger D}\right)=\int d^4x\mbox{tr}
   <x|e^{-tD^\dagger D}|x>.
\end{equation}
The plane wave basis $|p>$ significantly simplifies our calculations
and leads to the integral representation       
\begin{equation}
\label{logdet2}  
     W[Y]=\frac{1}{2}\int d^4x\int\frac{d^4p}{(2\pi )^4}
          \int^\infty_0\frac{dt}{t^3}\rho (t,\Lambda^2)
          e^{-p^2}\mbox{tr}\left(e^{-t({\cal M}^2+A)}\right)\cdot 1.
\end{equation}
Here $A=B-2ip\partial /\sqrt{t}$, and the trace is calculated in 
flavor space.
 
%%%%%%%%%%%%%%%%%%%%%%%%%%%%%%%%%%%%%%%%%%%%%%%%%%%%%%%%%%%%%%%%%%%%%%%%%%

\section{Asymptotic Schwinger -- DeWitt expansion}

%%%%%%%%%%%%%%%%%%%%%%%%%%%%%%%%%%%%%%%%%%%%%%%%%%%%%%%%%%%%%%%%%%%%%%%%%%
Before proceeding with our calculations this is the right place to say
several words about the standard asymptotic Schwinger -- DeWitt
expansion. If the mass matrix ${\cal M}$ has the degenerate form,
than $[{\cal M},A]=0$ and we find
\begin{equation}
\label{factor}
   \mbox{tr}\left(e^{-t({\cal M}^2+A)}\right)=
   e^{-tm^2}\mbox{tr}\left(e^{-tA}\right)=
   e^{-tm^2}\mbox{tr}\left(\sum^\infty_{n=0}t^n a_n\right).
\end{equation}
Here $a_n$ are the Seeley -- DeWitt coefficients which depend on 
background fields and their derivatives. The integration over momentum
and proper time in (\ref{logdet2}) can be readily done and we
obtain the well known result    
\begin{equation}
\label{logdet3}  
     W[Y]=\int\frac{d^4x}{32\pi^2}\sum_{n=0}^{\infty}
          J_{n-1}(m^2)\mbox{tr}(a_n),
\end{equation}
where integrals $J_n(m^2)$ are given by 
\begin{equation}
\label{Jn}
   J_n(m^2)=\int^\infty_0\frac{dt}{t^{2-n}}e^{-tm^2}\rho 
   (t,\Lambda^2).
\end{equation}
Independently on the type of regularization the following property 
is fulfilled 
\begin{equation}
\label{JnJ0}
   J_n(m^2)=\left(-\frac{\partial}{\partial m^2}\right)^nJ_0(m^2).
\end{equation}
Choosing $\rho (t,\Lambda^2)=1-(1+t\Lambda^2)e^{-t\Lambda^2}$, which
corresponds to two subtractions, one finds 
\begin{equation}
\label{J0}
   J_0(m^2)=\Lambda^2-m^2\ln\left(1+\frac{\Lambda^2}{m^2}\right).
\end{equation}
We see that functions $J_n(m^2)$, starting from $n>1$, have the 
asymptotic behavior $J_n(m^2)\sim m^{-2(n-1)}$ at large values 
of $m^2$, i.e. we obtain the inverse mass expansion. To warrant 
the convergence of the series it is necessary not only that the 
mass $m$ be different from zero, but also that the background 
fields change slowly at distances of the order of the Compton 
wavelength $(1/m)$ of the fermion field. If these criteria are 
not fulfiled then the creation of real pairs gets essential and 
this expansion is not useful for calculations. 

The remarkable property of the considered expansion is the gauge
covariance of the Seeley -- DeWitt coefficients.

%%%%%%%%%%%%%%%%%%%%%%%%%%%%%%%%%%%%%%%%%%%%%%%%%%%%%%%%%%%%%%%%%%%%%%%%%

\section{Non-degenerate mass case}

%%%%%%%%%%%%%%%%%%%%%%%%%%%%%%%%%%%%%%%%%%%%%%%%%%%%%%%%%%%%%%%%%%%%%%%%%
Let us return back to the formula (\ref{logdet2}) and show the 
way how to develop the abovementioned tool for the case 
$[{\cal M},A]\neq0$. As a first step let us use the formula 
\begin{equation} 
   \mbox{tr}\left(e^{-t({\cal M}^2+A)}\right)=\mbox{tr}\left(
   e^{-t{\cal M}^2}\left[1+\sum_{n=1}^{\infty}(-1)^nf_n(t,A)
   \right]\right), 
\end{equation}
to factorize an exponent with the non commuting mass matrix 
${\cal M}$. Here
\begin{equation}
\label{fnA}
   f_n(t,A)=\int^t_0ds_1\int^{s_1}_0ds_2\ldots
   \int^{s_{n-1}}_0ds_n A(s_1)A(s_2)\ldots A(s_n),
\end{equation}
where $A(s)=e^{s{\cal M}^2}Ae^{-s{\cal M}^2}$. It is true, that
\begin{equation}
   \mbox{tr}\left[e^{-t{\cal M}^2}f_n(t,A)\right]=
   \frac{t^n}{n!}\sum_{i_1,i_2,\ldots i_n}^{N_f}c_{i_1i_2\ldots i_n}(t)
   \frac{1}{n}\sum_{perm}\mbox{tr}(A_{i_1}A_{i_2}\ldots A_{i_n}),  
\end{equation}
where $A_i=E_iA$ by definition, a second summation means that
$n$ possible cyclic permutations of operators inside the trace
must be done. The coefficients $c_{i_1i_2\ldots i_n}(t)$ are 
totally symmetric with respect to permutations of indices and
are calculated easily \cite{Osipov:2001}, for instance
$$
   c_i(t)=e^{-tm^2_i},\quad 
   c_{ij}(t)=\frac{e^{-tm_i^2}-e^{-tm_j^2}}{t\Delta_{ji}},
$$
$$ 
   c_{ijk}(t)=\frac{2}{t^2}\left(\frac{e^{-tm_i^2}}{\Delta_{ji}\Delta_{ki}}
             +\frac{e^{-tm_j^2}}{\Delta_{kj}\Delta_{ij}}
             +\frac{e^{-tm_k^2}}{\Delta_{ik}\Delta_{jk}}\right),
$$
\begin{equation}
\label{ci}
   c_{ijkl}=\frac{3!}{t^3}\left(
            \frac{e^{-tm_i^2}}{\Delta_{li}\Delta_{ki}\Delta_{ji}}+
            \frac{e^{-tm_j^2}}{\Delta_{ij}\Delta_{lj}\Delta_{kj}}+
            \frac{e^{-tm_k^2}}{\Delta_{jk}\Delta_{ik}\Delta_{lk}}+
            \frac{e^{-tm_l^2}}{\Delta_{kl}\Delta_{jl}\Delta_{il}}\right).
\end{equation}
Here $\Delta_{ij}\equiv m_i^2-m_j^2$. In the case of coincidence 
of indices one can get that $c_i=c_{ii}=c_{ii\ldots i}$.  

Therefore the heat kernel is represented as a sum, every term of which
contains the coefficient $c_{ij\ldots}(t)$, multiplyed by the 
corresponding trace from the product of operators $A_i$. The following
integration over $t$ replaces the $t$-dependent part in these terms
by the integrals $J_l(m_i^2)$ and we finally will face the following
problem: every term of this expansion will not be invariant with 
respect to the transformations of the chiral group, although the total 
effective action $W[Y]$ will possess this property. One needs the algorithm
which automatically groups the terms in chiral invariant blocks. This 
problem has been solved in the papers \cite{Osipov:2001} 
on the basis of recurrence relations
\begin{equation}
\label{recfor}
   J_l(m_j^2)-J_l(m^2_i)=\sum^\infty_{n=1}\frac{\Delta_{ij}^n}{2^nn!}
   \left[J_{l+n}(m_i^2)-(-1)^nJ_{l+n}(m_j^2)\right].
\end{equation}
There it has been shown that the essence of the problem is confined
to the correct choosing of the factorized combination, built from the
functions $J_i(m_j^2)$, namely, the effective action $W[Y]$ must be
represented in the form 
\begin{equation}
\label{logdet5} 
   W[Y]=\int\frac{d^4x}{32\pi^2}\sum^\infty_{i=0}I_{i-1}\mbox{tr}(b_i),
   \quad I_i\equiv\frac{1}{N_f}\sum_{j=1}^{N_f}J_i(m^2_j).
\end{equation}
In this case the background fields are {\it automatically} combined
in the covariant coefficients $b_i$. For example, for $N_f=3$ the
first four coefficients are  
\begin{eqnarray}
\label{coeff}
     b_0&=&1, \quad b_1=-Y, \quad b_2=\frac{Y^2}{2}
         +\frac{\Delta_{12}}{2}\lambda_3Y+\frac{1}{2\sqrt{3}}
         \left(\Delta_{13}+\Delta_{23}\right)\lambda_8Y,
         \nonumber \\
     b_3&=&-\frac{Y^3}{3!}
           -\frac{1}{12}(\partial Y)^2 
           -\frac{1}{12}\Delta_{12}
           \left(\Delta_{31}+\Delta_{32}\right)\lambda_3Y
           \nonumber\\         
        &+&\frac{1}{12\sqrt{3}}\left[\Delta_{13}(\Delta_{21}+\Delta_{23})
           +\Delta_{23}(\Delta_{12}+\Delta_{13})\right]\lambda_8Y
           \nonumber\\
        &+&\frac{1}{4\sqrt{3}}\left(\Delta_{31}+\Delta_{32}\right)
           \lambda_8Y^2
           +\frac{1}{4}\Delta_{21}\lambda_3Y^2.
\end{eqnarray}

Several comments are in order here. First of all it is obvious
that in the limiting case of equal masses $m_1=m_2=\ldots =m_{N_f}$ 
our result coincides with the standard Schwinger -- DeWitt expansion.
If the masses are not equal, the series (\ref{logdet5}) is a
generalization of the well-known result (\ref{logdet3}). Instead
of the asymptotic Seeley -- DeWitt coefficients $a_n$ come the 
coefficients $b_n$. Indeed one can check that if the operator 
$D^\dagger D$ transforms in the adjoint representation
$\delta (D^\dagger D)=i[\omega,D^\dagger D]$, then $b_n$ are also
covariant, i.e. $\delta b_n=i[\omega,b_n]$, where
$\omega =\alpha +\gamma_5\beta$ are the parameters of the global
infinitesimal chiral transformations. It also should be mentioned
that, rigorously speaking, the obtained expansion is not an exact 
inverse mass expansion. Although the integrals $I_l$ for $l\ge 1$ 
possess the necessary form for asymptotic behaviour $I_{l+1}(m_i^2)\sim 
m_i^{-2l}$, the coefficients $b_l$ depend on the differences of
masses, which may influence the character of the expansion.
However general symmetry requirements are a more serious and 
stringent argument in favour of the obtained series as compared to
the result of the thorough study \cite{Lee:1989} based only on the 
idea of a $1/m^2$ expansion.

%%%%%%%%%%%%%%%%%%%%%%%%%%%%%%%%%%%%%%%%%%%%%%%%%%%%%%%%%%%%%%%%%%%%%%%%%

\section{Relation between coefficients $b_n$ and $a_n$} 
 
%%%%%%%%%%%%%%%%%%%%%%%%%%%%%%%%%%%%%%%%%%%%%%%%%%%%%%%%%%%%%%%%%%%%%%%%%
The problem of the calculation of the generalized heat kernel
coefficients is a more complicated mathematical problem than
the calculation of the standard Seeley -- DeWitt coefficients.
However one can significantly simplify this problem
\cite{Salcedo:2001}, if from the very beginning one uses
the transformation properties of the coefficients $b_n$. 
Indeed, let us return back to the formula (\ref{factor}) and 
extend it to the case $[{\cal M},Y]\neq 0$, omitting for simplicity
all terms with derivatives in $A$. It is remarkable that already at
this stage one can take into account the two main conclusions which
we have found in the previous section: the form of the factorized
part depending on the mass and gauge covariance of coefficients $b_n$. 
The first aim can be reached through the definition 
\begin{equation}
\label{bra}
   <A>\equiv\frac{\mbox{tr}(A)}{\mbox{tr}(1)},\quad 
   {\bar{\cal M}}^2={\cal M}^2-<{\cal M}^2>,
\end{equation}
the second one through putting
\begin{equation}
\label{covbk}
   \left(\bar{\cal M}^2+Y\right)^n\equiv
   \sum_{k=0}^n(-1)^kC^k_n<\left(\bar{\cal M}^2\right)^{n-k}>b_k.
\end{equation}
Since the left side of Eq.(\ref{covbk}) transforms covariantly,
i.e. $(\bar{\cal M}^2+Y)^\Omega =\Omega^{-1}(\bar{\cal M}^2+Y)\Omega$,
and on the right side the term $<\bar{\cal M}^2>$ is 
invariant with respect to the chiral transformations $\Omega$, it is
obvious, that $b_k(Y^\Omega ,\bar{\cal M}^{2\Omega})
=\Omega^{-1}b_k(Y,\bar{\cal M}^2)\Omega$. To see how these 
definitions work let us consider the exponent in the heat kernel
(\ref{factor})
\begin{eqnarray}
   &&e^{-t({\cal M}^2+Y)}=e^{-t<{\cal M}^2>}\sum_{n=0}^\infty
   \frac{(-t)^n}{n!}\left(\bar{\cal M}^2+Y\right)^n \nonumber \\
   &=&e^{-t<{\cal M}^2>}<e^{-t\bar{\cal M}^2}>\sum_{n=0}^\infty
   \frac{t^n}{n!}b_n=<e^{-t{\cal M}^2}>\sum_{n=0}^\infty
   \frac{t^n}{n!}b_n.
\end{eqnarray}
Since 
\begin{equation}
   <e^{-t{\cal M}^2}>=\sum_{i=1}^{N_f}e^{-tm_i^2}<E_i>=\frac{1}{N_f}
                      \sum_{i=1}^{N_f}e^{-tm_i^2},
\end{equation} 
the following integration over $t$ leads us to the known result
(\ref{logdet5}), and the coefficients $b_n$ can be found using
the definition (\ref{covbk}). The same definition allows us
to relate $b_n$ with the linear combination of the standard
Seeley -- DeWitt coefficients $a_i(Y)$, where the replacement
$Y\rightarrow Y+\bar{\cal M}^2$ should be done \cite{Salcedo:2001}
\begin{equation}
   b_n=\sum_{i=0}^{n}\frac{\beta_{n-i}}{(n-i)!}a_i(Y\rightarrow 
       Y+\bar{\cal M}^2).
\end{equation}
The parameters $\beta_i$ depend only on the masses of fermion 
fields. To establish the form of this dependence it is sufficient
to calculate $b_n$ in the simplest case, when in the elliptic
operator under consideration all terms with derivatives are
omitted. This problem is much simpler to solve than the calculation 
of the coefficients $b_n$ from the very beginning.
 
%%%%%%%%%%%%%%%%%%%%%%%%%%%%%%%%%%%%%%%%%%%%%%%%%%%%%%%%%%%%%%%%%%%%

\section{ACKNOWLEDGMENTS}

%%%%%%%%%%%%%%%%%%%%%%%%%%%%%%%%%%%%%%%%%%%%%%%%%%%%%%%%%%%%%%%%%%%%
This work is supported by grants provided by Funda\c{c}\~ao para
a Ci\^encia e a Tecnologia, POCTI/2000/FIS/35304 and 
NATO ``Outreach'' Cooperation Program.

%%%%%%%%%%%%%%%%%% REFERENCES %%%%%%%%%%%%%%%%%%%%%%%%%%%%%%%%%%%%%%%%%%%%%%
\baselineskip 12pt plus 2pt minus 2pt

%%%%%%%%%%%%%%%%%%%%%%%
\end{document}